\documentclass[aps,prb,preprint,groupedaddress]{revtex4-1}
\usepackage{hyperref}
\usepackage{amssymb}
\usepackage{graphicx, amsmath, amsfonts,ae}

\newcommand{\ud}[0]{d}                           % Defines an integration d
\newcommand{\op}[1]{\hat{#1}}                    % Defines operator symbol
\newcommand{\mat}[1]{\overleftrightarrow{#1}}    % Defines matrix symbol
\newcommand{\eq}[1]{Eq. (\ref{eq:#1})}           % defines eqation notation
\newcommand{\tab}[1]{Tab. \ref{tab:#1}}          % defines table
\newcommand{\fig}[1]{Fig. \ref{fig:#1}}          % Defines figure notation
\begin{document}
\title{Correlation and dimensional effects of trions in carbon nanotubes}
\author{Troels F. Rønnow} 
\email{tfr@nanophysics.dk}
\homepage{http://www.nanophysics.dk}
\author{Thomas G. Pedersen}
\email{tgp@nano.aau.dk}
\affiliation{Department of Physics and Nanotechnology, Aalborg
  University, Skjernvej 4A, 9220 Aalborg Øst, Denmark}
\author{Horia D. Cornean}
\email{cornean@math.aau.dk}
\affiliation{Department of Mathematical Sciences, Aalborg University,
  Frederik Bajers Vej 7G, 9220 Aalborg Øst, Denmark}

\begin{abstract}
We study the binding energies of singlet trions, i.e. charged excitons,
in carbon nanotubes. The problem is modeled,
through the effective-mass model,
as a three-particle complex
on the surface of a cylinder, which we investigate using both one- and
two-dimensional expansions of the wave function. The effects of
dimensionality and correlation are studied in detail. We find that the
Hartree-Fock approximation significantly underestimates the trion
binding energy. Combined with band structures calculated using 
a non-orthogonal nearest neighbour 
tight binding model, the results from the cylinder
model are used to compute physical binding energies for a wide selection
of carbon nanotubes. In addition, the dependence on dielectric screening
is examined. Our findings indicate that trions are detectable at room
temperature in carbon nanotubes with radius below $8$\AA{}.
\end{abstract}
\pacs{71.35.Pq}

\maketitle
\section{Introduction}
Many applications of semiconducting carbon nanotubes (CNTs) rely on the
formation of junctions using either doped \cite{citeulike:3887558} or
asymmetrically gated CNTs \cite{citeulike:6498581}. Such junctions are
the basis for e.g. CNT based light emitting diodes
\cite{citeulike:3887558,citeulike:3282181,citeulike:3471809} and other
optoelectronic devices. In the description of the optical properties of
CNTs, excitons are known to significantly influence emission
\cite{citeulike:6603030,citeulike:6598379} and absorption
\cite{citeulike:4014877}. The importance of excitons in CNTs is a
consequence of the extremely large binding energy of several hundred meV
\cite{citeulike:6598379,citeulike:6603030,citeulike:4014877}. Thus, in 
an optically excited junction with injected 
carriers, an exciton in combination with either a hole or an electron
may form a three-particle complex known as a trion. 
It has been shown experimentally that trions form in biased GaAs quantum wells\cite{citeulike:6877050} under optical excitation via the interaction between excitons and injected electrons and holes. Consequently, similar processes are expected to occur in CNTs. 
Hence, if the binding
energy is sufficiently large, these complexes 
are expected to be 
of importance for
optoelectronic devices and could even be detectable at room
temperature.\\

In traditional semiconductors, trion binding energies and optical
spectra have been studied in one
\cite{citeulike:6056219,citeulike:6056170}, two
\cite{citeulike:3880291,citeulike:3880278,citeulike:3852371,citeulike:3852347,citeulike:5971241}
and three dimensions \cite{citeulike:5971241}. In contrast, trions in
CNTs have only been studied in an effectively
one-dimensional model
\cite{troelsfr2009}, in which carriers were assumed to be completely
delocalised around the circumference of the tube. Within this model, it
was found that singlet and triplet trions in CNTs are stable against
dissociation into excitons and free carriers. Furthermore, the difference
in energy between the positive $S^+$ and negative $S^-$ singlet trion
states was rather small for mass fractions $\sigma \equiv m_e/m_h \in
[0.8;1]$, where $m_e$ is the effective electron mass and $m_h$ is the
effective hole mass. Also, numerical trion binding energies were
estimated for CNTs with radius from 2.5{\AA} to 10{\AA}. It is clear,
however, that a one-dimensional model may have limited applicability for
large diameter CNTs, in which motion around the circumference is an
important degree of freedom. \\ 

The one-dimensional model used in Ref. \onlinecite{troelsfr2009} has
many potential applications as it greatly simplifies the complexity of
the calculations for trions, biexcitons, charged biexcitons,
etc. However, it still remains to assess the quality of this
approximation and, therefore, it is important to compare one- and
two-dimensional solutions for trions. 
It is clear that the effective-mass model used in Ref. \onlinecite{troelsfr2009} is an 
approximation to more accurate \textit{ab initio} approaches. However, comparison with \textit{ab initio} results for the exciton energy and its scaling with tube radius\cite{citeulike:2453523} as well as the exciton wave function \cite{citeulike:6565076} demonstrates excellent agreement between these different approaches\cite{citeulike:2453523,citeulike:6565076}. 
In this paper, we investigate two
simplifications of the full 
effective-mass 
trion problem:  
First, the one-dimensional
model is examined by comparison to two-dimensional 
solutions of the full problem. 
Second, the accuracy of the Hartree-Fock approximation for
trion binding energies is determined in both one- and two-dimensional
models. As our starting point for both models, we expand the wave 
function in a convenient basis. 
The same basis expansion is applied in the Hartree-Fock calculation,
which is briefly outlined.
Next, we investigate the binding energies of trions, in both
the one- and two-dimensional models, as functions of cylinder radius
$r$ and mass fraction $\sigma$. We compare the Hartree-Fock solutions
with the full solutions (at $\sigma=0.0$), and we analyze the
distribution of electrons along the circumference in a negative trion
state $S^-$ for the two methods. Finally, the CNT band structure is used
to convert model results into physical binding energies for a wide range
of CNTs, using both the one- and two-dimensional models. 

\section{Theory}
\label{sec:theory}
\begin{figure}
  \centering
  \includegraphics[width=0.5\textwidth]{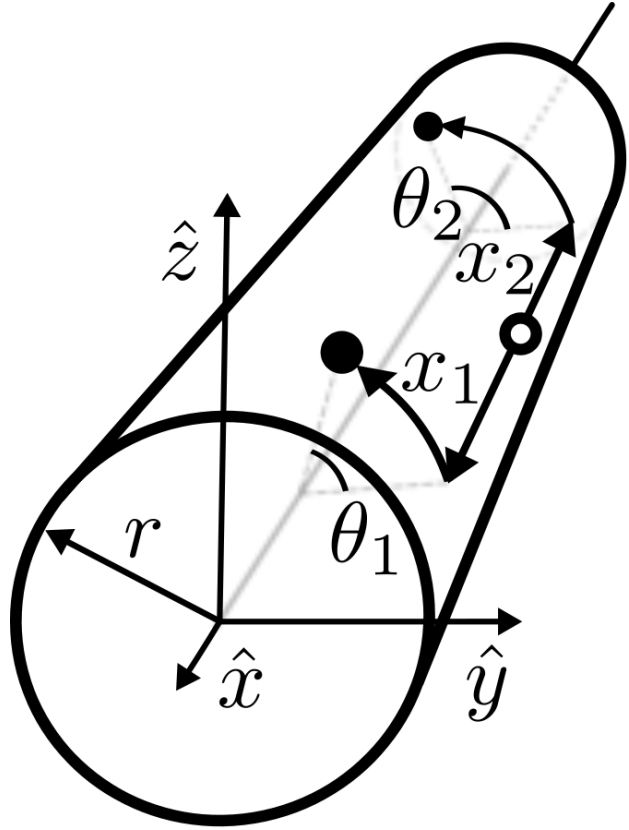}% 
  \caption{Geometry of a negative trion on the surface of a cylinder of
    radius $r$, described in relative coordinates. The filled circles
    illustrate electrons and the open circle illustrates a hole.}
  \label{fig:geometry}
\end{figure}

Adopting the fundamental trion equation from
Ref. \onlinecite{citeulike:5971241} and introducing relative cylindrical
coordinates along with removal of center-of-mass motion, we get the
Hamiltonian for the negative trion
\begin{equation}
  \label{eq:operator_minus}
  \op{H}_{-} = \op{h}_1 +\op{h}_2 -\frac{2\sigma}{1+\sigma}\left(\frac{\partial^2 }{\partial x_1\partial x_2}+\frac{1}{r^2}\frac{\partial^2 }{\partial \theta_1\partial \theta_2}\right) + V(x_1-x_2, \theta_1-\theta_2) ,
\end{equation}
where $r$ is the nanotube radius and $x_i$ and $r\theta_i$ denote axial
and circumference coordinates, respectively, as shown in \fig{geometry}.
The exciton operators, $\op{h}_1$ and $\op{h}_2$, are given by
\begin{equation}
  \op{h}_i = -\frac{\partial^2 }{\partial x_i^2}-\frac{1}{r^2}\frac{\partial^2 }{\partial \theta_i^2}-V(x_i, \theta_i),
\end{equation}
and the Coulomb potential in the cylinder geometry is given as
\cite{citeulike:3870975}
\begin{equation}
  V(x,\theta) = \frac{2}{\sqrt{x^2+4r^2\sin^2\frac{\theta}{2}}}.
\end{equation}
Here, the coordinate pairs $(x_1,\theta_1)$ and $(x_2,\theta_2)$
describe the motion of the first and second electron relative to the
hole, respectively. The positive trion Hamiltonian $\op{H}_{+}$ is found
by the replacement $\sigma \rightarrow \sigma^{-1}$. Note that
$\op{H}_{-}$ and $\op{H}_{+}$ have incorrectly been interchanged in
Ref. \onlinecite{troelsfr2009}.  
The Hamiltonian is expressed in natural exciton units, i.e. effective
Bohr radii $a_B^*$ and effective Rydbergs $Ry^*$ for distances and
energies, respectively. \\

If the radius $r$ is much smaller than the effective Bohr radius
$a_B^*$, the electrons are highly delocalised around the circumference
and the angular dependence of the wave function for the lowest states will
be nearly constant. In contrast, when $r$ is comparable to $a_B^*$, the
angular dependence is expected to have a cusp-like behaviour
similar to the one found for excitons\cite{citeulike:3870975}. A
function that suitably describes this behaviour is the absolute value of
sine $|\sin \frac{\theta}{2} |$. 
Thus, it would be reasonable to model the angular part of the wave
function as a linear combination of the two limits. Moreover, to model
the electron-electron repulsion for $\op{H}_{-}$ and hole-hole repulsion
for $\op{H}_{+}$, the function $|\sin(\frac{\theta_1-\theta_2}{2})|$ constitutes a
sensible choice as it attains its minimum along the diagonal
$\theta_1=\theta_2$. 
Having this in mind, we expand the wave function according to 
\begin{equation}
\psi(x_1, \theta_1, x_2,\theta_2 ) = \sum_{i,j,k}^{N_1,N_2,N_3}\sum_{l}^{4}c_{ijkl}\phi_{i}(x_1)\phi_{j}(x_2)\phi_{k}(x_1-x_2)\varphi_l(\theta_1, \theta_2),
\end{equation}
with Gaussians $\phi_i(x) = e^{-\alpha_i x^2}$ along the translational
direction and an angular basis $\{\phi_l\}_{l=1}^4$ along the
circumference, where
\begin{equation}
  \varphi_l(\theta_1, \theta_2) =\frac{1}{2\pi} \left\{
      \begin{array}{cl}
        1 & l=1,\\
        |\sin\frac{\theta_1}{2}| & l=2,\\
        |\sin\frac{\theta_2}{2}| & l=3,\\
        |\sin\frac{\theta_1-\theta_2}{2}| & l=4.
      \end{array}
\right.
\end{equation}
Here, $i$, $j$ and $k$ run from $1$ to $N_1$, $N_2$ and $N_3$,
respectively, and $N = N_1\cdot N_2\cdot N_3$ is the total number of
Gaussians used in the expansion. The expansion coefficients $c_{ijkl}$
are obtained from the eigenvector $\vec{c}$ of a matrix equation of the
form $(\mat{K}+\mat{U})\cdot \vec{c}=E_T\mat{S}\cdot \vec{c}$. We 
define the trion binding energy
$E_B\equiv E_X-E_T$ as the difference between the exciton binding
energy $E_X$ and the trion energy $E_T$, where a positive $E_B$
indicates a stable trion. The kinetic matrix $\mat{K}$ and overlap
matrix $\mat{S}$  elements turn out as $S_{ijkl,i'j'k'l'} =
S^{(T)}_{ijk,i'j'k'}S^{(C)}_{ll'}$ and
\begin{multline}
  K_{ijkl,i'j'k'l'}=S^{(T)}_{ijk,i'j'k'}\left(K_{ll'}^{(C)}+\frac{2\sigma}{1+\sigma}K_{ll'}^{(CM)}\right) \\
  + \left(K^{(T)}_{ijk,i'j'k'} + K^{(T)}_{jik,j'i'k'}+\frac{2\sigma}{1+\sigma}K_{ijk,i'j'k'}^{(TM)} \right)S_{ll'}^{(C)}
\end{multline}
with
\begin{eqnarray*}
  S_{ijk,i'j'k'}^{(T)} &=&\frac{\pi}{\sqrt{(\alpha_i+\alpha_{i'})(\alpha_j+\alpha_{j'})(\alpha_k+\alpha_{k'})}},\\
  K^{(T)}_{ijk,i'j'k'} &=&  
 \frac{
2 \pi  
(
  \alpha_{k'}\alpha_k (\alpha_j+\alpha_{j'}+\alpha_i+\alpha_{i'})
+ \alpha_{i'}\alpha_i(\alpha_{k'}+\alpha_{k})
)
}
{
((\alpha_j+\alpha_{j'})
   (\alpha_i+\alpha_{i'}+\alpha_k+\alpha_{k'})
   +(\alpha_i+\alpha_{i'})
   (\alpha_k+\alpha_{k'}))^{3/2}
}\nonumber \\
&+& \frac{
2 \pi  
  (\alpha_j+\alpha_{j'})(\alpha_{k'}\alpha_i+\alpha_{i'}\alpha_k+\alpha_i \alpha_{i'})
}
{
((\alpha_j+\alpha_{j'})
   (\alpha_i+\alpha_{i'}+\alpha_k+\alpha_{k'})
   +(\alpha_i+\alpha_{i'})
   (\alpha_k+\alpha_{k'}))^{3/2}
},\\
K^{(C)}_{ll'} &=& \frac{1}{8}(1-\delta_{1l})\delta_{ll'},\\
K^{(TM)}_{ijk,i'j'k'} &=&  -
\frac{2 \pi  (\alpha_k \alpha_{k'} (\alpha_i+\alpha_{i'}+\alpha_j+\alpha_{j'})
  +\alpha_i \alpha_j
   \alpha_{k'}+\alpha_{i'} \alpha_{j'} \alpha_k)}
{((\alpha_i+\alpha_{i'})
   (\alpha_j+\alpha_{j'}+\alpha_k+\alpha_{k'})+(\alpha_j+\alpha_{j'})
   (\alpha_k+\alpha_{k'}))^{3/2}},\\
K^{(CM)}_{ll'} &=&  -\frac{1}{8}\delta_{4l}\delta_{ll'},
\end{eqnarray*}
where $\delta_{ij}$ is the Kronecker delta and with the angular overlap
integrals $S_{ll'}^{(C)}$ given in \tab{overlap}.
\begin{table}
  \centering
  \begin{tabular}{c|cccc}
    $S_{ll'}^{(C)}$ & 1 & 2  & 3 & 4 \\\hline
    1 & 1               & $\frac{2}{\pi}$     & $\frac{2}{\pi}$   & $\frac{2}{\pi}$\\
2 & $\frac{2}{\pi}$ & $\frac{1}{2}$ & $\frac{4}{\pi^2}$ &
$\frac{4}{\pi^2}$\\ 3 & $\frac{2}{\pi}$ & $\frac{4}{\pi^2}$ &
$\frac{1}{2}$ & $\frac{4}{\pi^2}$\\ 4 & $\frac{2}{\pi}$ &
$\frac{4}{\pi^2}$ & $\frac{4}{\pi^2}$ & $\frac{1}{2}$
  \end{tabular}
  \caption{Table of angular overlap integrals.}
  \label{tab:overlap}
\end{table}
In the derivation, care should be taken as the second derivative of the
absolute value of a sine gives a delta function, which takes many of the
kinetic energy elements to zero. The potential energy matrix elements
turn out to be slightly more complicated and can be expressed in terms
of 4 different Meijer {\em G} functions.\\ 
In order to formulate the problem in terms of the Hartree-Fock
approximation we now assume that the eigenfunction of the operator
\eq{operator_minus} is a Slater determinant of single electron states
$\chi(x, \theta)$ with anti-symmetry in the spin part. From standard
quantum mechanics \cite{QUANTUMWORLD} we then obtain the Fock operator
for the corresponding system
\begin{equation} 
  \op{F} = - \frac{\partial^2}{\partial x^2}-
  \frac{1}{r^2}\frac{\partial^2}{\partial \theta^2} - \frac{2}{\sqrt{x^2
      +4r^2\sin^2\frac{\theta}{2} }}+V_H(x,\theta),
\end{equation} 
where the Hartree potential $V_H$ is given by
\begin{equation} 
  V_H(x,\theta)=
  2\int_{-\infty}^{\infty}\ud x'
  \int_{-\pi}^{\pi} \ud \theta'\frac{|\chi(x',\theta')|^2}{\sqrt{(x-x')^2
      +4r^2\sin^2\frac{\theta-\theta'}{2} }}.
\end{equation} 
From the eigenvalue equation $\op{F}\chi(x,\theta) = \varepsilon \chi(x,
\theta)$ the Hartree-Fock eigenvalues are found and the lowest one
$\varepsilon_0$ is related to the trion ground state energy in the usual way 
\begin{equation} 
  E^{(HF)}_T = 2\varepsilon_0-\int_{-\infty}^{\infty}\ud x
  \int_{-\pi}^{\pi} \ud \theta |\chi(x,\theta)|^2V_H(x,\theta).
\end{equation} 
Expanding $\chi(x, \theta)$ in a basis of Gaussians along the
translational direction and absolute value of sines along the
circumference, as done for the full problem, we are capable of solving
the Hartree-Fock eigenvalue equation self-consistently.

\section{Results and discussion} 
Optimisation of the Gaussian coefficients was carried out using a
steepest descent method. The following coefficients, in units of
$a_B^{*-2}$, were found for the different optimisations: $\alpha_i \in
\{$$0.143,$$1.16,$$4.98,$$29.0,$$250\}$ for both the 1D and 2D exciton
and $\alpha_i, \alpha_j, \alpha_k \in
\{$$0.0651,$$0.145,$$1.68,$$9.65,$$48.7\}$ for the 1D trion. For the 2D
trion, the Gaussian coefficients were found to be $\alpha_i, \alpha_j
\in \{$$0.165,$$1.68,$$9.65,$$48.7\}$ for the functions
$e^{-\alpha_ix^2}$ and $e^{-\alpha_jy^2}$, and $\alpha_k \in
\{$$0.0000171,$$1.68,$$9.98,$$48.7\}$ for the functions
$e^{-\alpha_k(x-y)^2}$. Similarly, the coefficients for the Hartree-Fock
calculations were found to be $\alpha_i,\alpha_j\in\{0.0648, 0.195,
1.04, 5.28, 27.5, 99.3, 250\}$ for both the 1D and 2D cases. All
optimisations were carried out at $r_0=0.1a_B^*$ and all coefficients
were subsequently scaled with $r_0^2/r^2$ for calculations involving
other radii.\\ 

\begin{figure}
  \centering
  \includegraphics[width=0.75\textwidth]{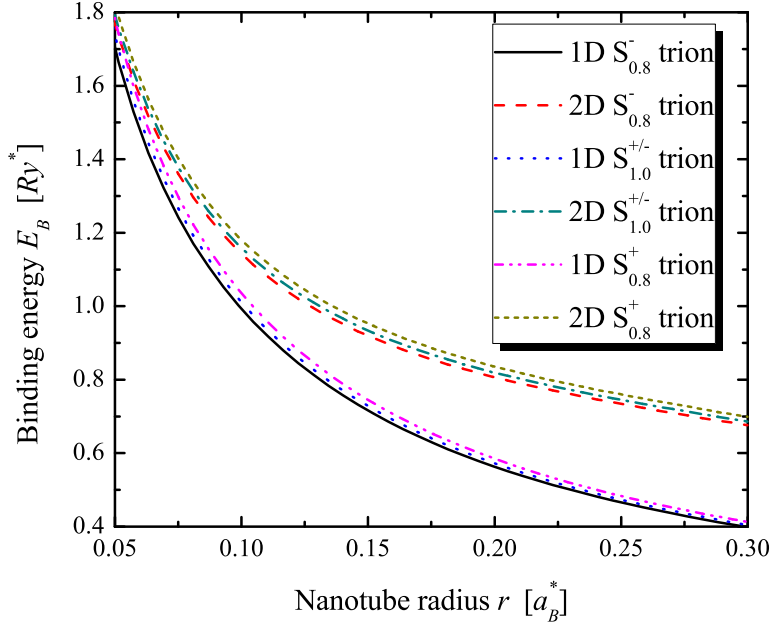}
  \caption{Comparison between the one- and two-dimensional models. For
    each of the two models three cases are shown, namely, $S^+_{0.8}$,
    $S^{+/-}_{1.0}$ and $S^-_{0.8}$. The positive and negative trion
    binding energies were calculated for $\sigma=0.80$. The
    $S^{+/-}_{1.0}$ binding energies were calculated for $\sigma=1.0$.}
  \label{fig:1vs2}
\end{figure}

The binding energies $E_B$ of trions were found as functions of radius
for both the one- and two-dimensional models. In \fig{1vs2}, it is seen
that the one- 
and two-dimensional results deviate less as $r$ approaches zero, as
expected. It is important to realize that the effective Bohr radius
$a_B^*$ is dependent on nanotube species via the effective masses and
dielectric constant $\varepsilon$. Due to the scaling of masses with
nanotube radius, it turns out that $r$ is, in fact, roughly proportional
to $a_B^*$. For CNTs embedded in a medium with $\varepsilon=3.5$ it is
found that\cite{citeulike:3870975} $r\approx 0.1a_B^*$, and for this
value the correction is roughly 13\% of the two-dimensional result. The
one-dimensional model performs significantly worse when the radius is
increased. For $r\approx 0.3a_B^*$ the one-dimensional result is 42\%
below the two-dimensional one and, as would be expected, the
one-dimensional model can only be considered a good approximation for
$r$ close to zero. Realistically, though, $r > 0.2a_B^*$ will only occur
for CNTs embedded in media with small screening $\varepsilon<1.75$.\\ 

\begin{figure}
  \centering
    \includegraphics[width=0.75\textwidth]{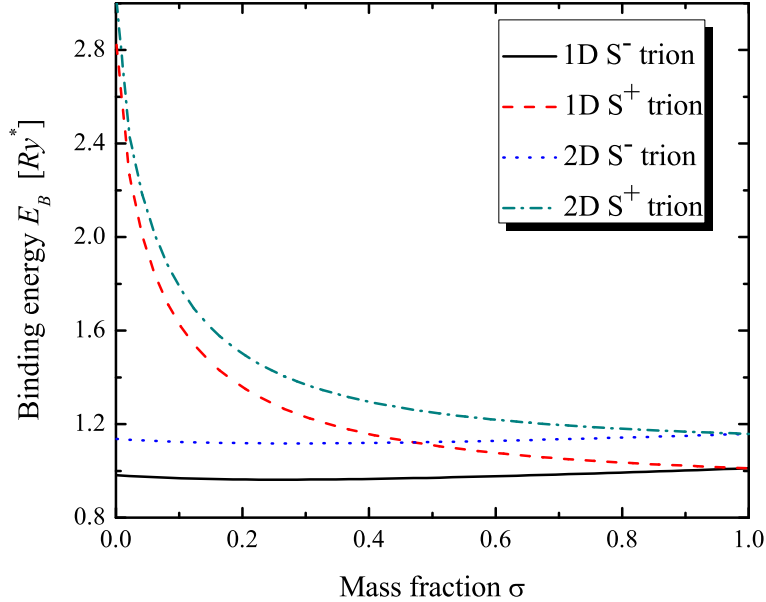}
    \caption{Trion binding energy as a function of the mass fraction
      $\sigma=m_e/m_h$ for the one- and two-dimensional models. The
      calculations were done for $r=0.1a_B^*$.}
  \label{fig:massfraction}
\end{figure}

From \eq{operator_minus} it follows that the mass fraction $\sigma$ affects the binding
energy via a mixed kinetic energy term coupling the two relative coordinates
$(x_1,\theta_1)$ and $(x_2,\theta_2)$. 
The mass fraction $\sigma$ of CNTs varies between 0.86 and 1.0
\cite{troelsfr2009} and to gain an idea of the upper and lower bounds of
the trion binding energies we have plotted the positive and negative
trions at $\sigma=0.8$ in \fig{1vs2}. Actual binding energies should lie
between the $S_{0.8}^-$ and $S_{1.0}^{+/-}$ energy curves and the
$S_{1.0}^{+/-}$ and $S_{0.8}^+$ curves for the negative and positive
trion, respectively. It is seen that the difference is fairly small for
both the one- and two-dimensional models. In \fig{massfraction}, we illustrate
the binding energies as functions of the mass fraction $\sigma$ for
both models for $r=0.1a_B^*$. In both
cases, the 
binding energy for the negative trion $S^-$ was found to have little
variation for $\sigma\in[0;1]$. As a consequence, the results calculated
for $S_{0.0}^{-}$ are, in fact, excellent approximations for both 
$S^{+}_{\sigma}$ and $S^{-}_{\sigma}$ with mass fractions in the range 
$\sigma \in [0.8;1]$. It follows that neglecting the mixed kinetic energy term 
is an acceptable approximation, which in the
worst case would give an error of 3.2\% at $r=0.1a_B^*$.\\ 

\begin{figure}
  \centering
  \includegraphics[width=0.75\textwidth]{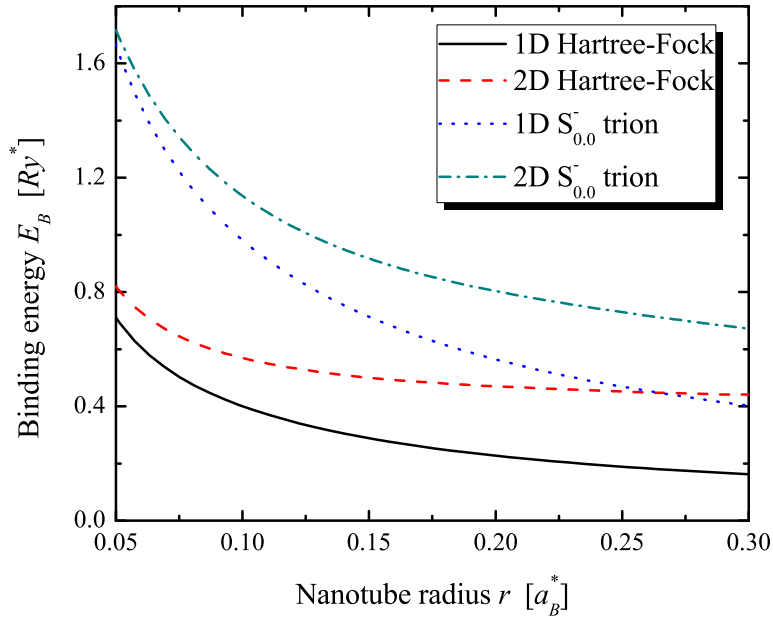}
  \caption{Solutions for the one- and two-dimensional models
    compared with the corresponding Hartree-Fock solutions. The
    calculations were carried out at $\sigma=0.0$.}
  \label{fig:HFvsfull}
\end{figure}

Next we calculated the binding energies using the Hartree-Fock
approximation. As demonstrated above, the mixed kinetic energy term is only
a minor correction for actual nanotubes. Thus, in the following we ignore this 
term and limit the discussion to 
$S_{0.0}^{-}$ states. In \fig{HFvsfull}, the Hartree-Fock energies are plotted
along with the $S^-_{0.0}$ trion energies for both one- and
two-dimensional solutions. The Hartree-Fock solutions significantly underestimate
trion binding energies is both cases. Differences up to a factor of 2,
in the two-dimensional case, were found. The best Hartree-Fock result
obtained in the interval $r\in[0;0.3]$ was at the end point $r=0.3a_B^*$
where the Hartree-Fock energy is 60\% of the correct result. 
The trion problem of two electrons interacting with a positive hole is mathematically similar to the He atom. 
Hence, the dramatic failure of the Hartree-Fock approximation is surprising at first sight. However,
the failure is readily ascribed to several characteristic factors in the trion case. 
Primarily, we focus here on the binding energy $E_B = E_X-E_T$ given as the difference 
between exciton and trion energies. Hence, relatively small errors in the Hartree-Fock 
trion energies $E_T^{(HF)}$ lead to large errors in the binding energies provided $E_T^{(HF)}-E_T$ 
is comparable to $E_X-E_T$. Secondly, \fig{HFvsfull} shows that the error increases
as we decrease the radius, i.e. when the effective dimension of the system
decreases from two towards unity. Actual CNTs can be regarded as effectively
1.7 dimensional systems\cite{tgpedersen2007}. The approximate treatment of electron-electron repulsion 
in the Hartree-Fock calculation is more serious in a low-dimensional geometry due to the increased
overlap between electrons enforced by the geometry. Hence, the Hartree-Fock error for the 
(three dimensional) He atom is small even if the energy is measured relative to the hydrogen atom.  
Finally, in the present problem, the strengths (charges) of the electron-hole and electron-electron are
identical. In He, the larger nuclear charge $Z=2$ enhances the electron-nucleus interaction. As a
result, the contribution to the energy from the electron-electron
repulsion is relatively smaller and, thus, the Hartree-Fock
approximation is less inaccurate.\\ 

\begin{figure}
  \centering
  \includegraphics[width=0.75\textwidth]{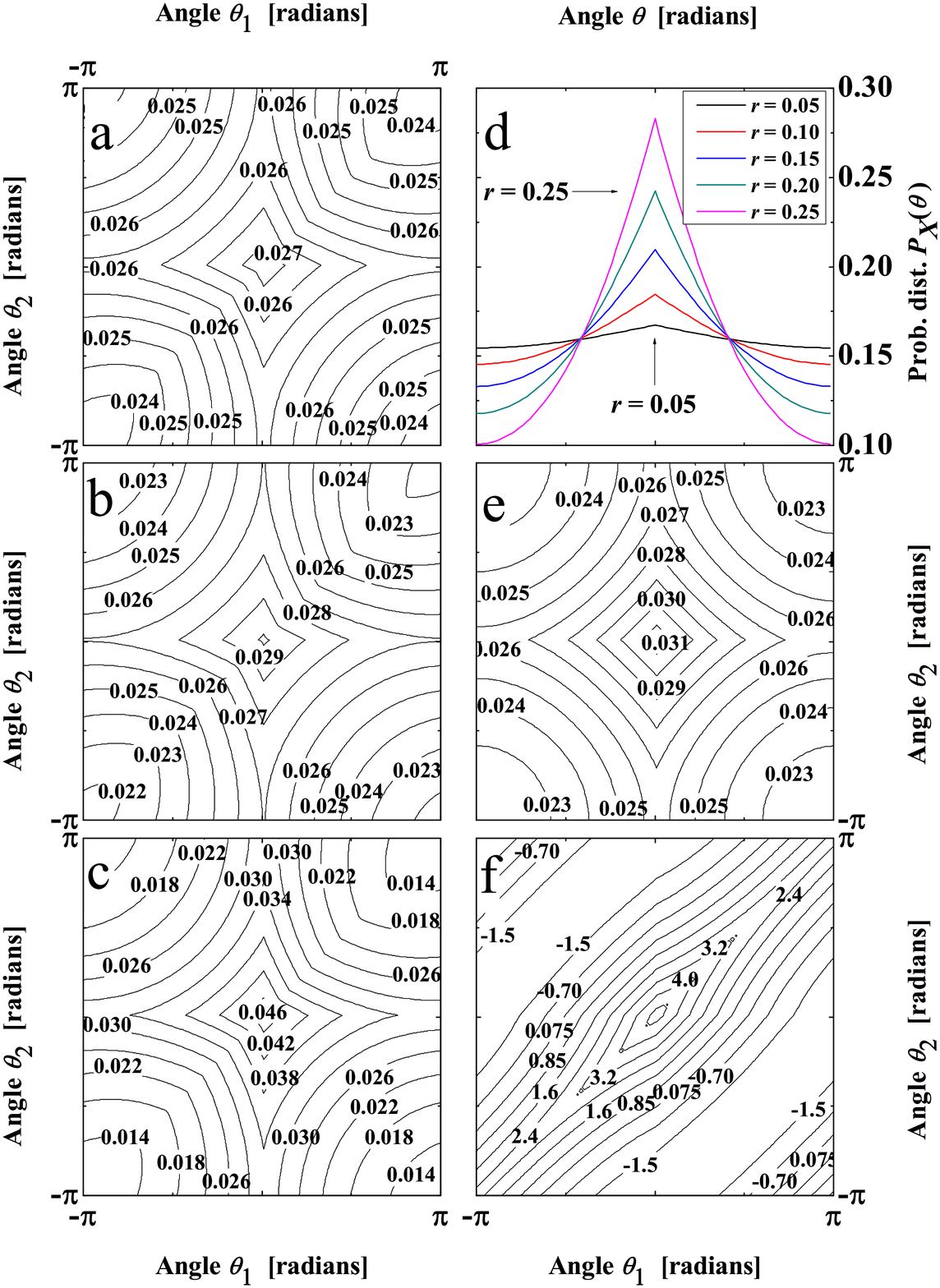}
  \caption{Probability distributions as a functions of angles for (a)
    $r=0.05a_B^*$, (b) $r=0.10a_B^*$ and (c) $r=0.25a_B^*$ for the
    singlet trion $S^{+/-}_{1.0}$ calculated with a basis expansion. The
    exciton probability distributions have been shown in (d) for the
    radius varying from $r=0.05a_B^*$ to $r=0.25a_B^*$. In (e), the
    Hartree-Fock probability distribution is shown at $r=0.1a_B^*$ and
    (f) shows percent wise difference between the Hartree-Fock solution
    and the solution obtained from the full problem at $r=0.1a_B^*$}
  \label{fig:wavefunction}
\end{figure}

To investigate the behaviour of the wave function around the
circumference the probability distribution, averaged over the
translational coordinates $x_1$ and $x_2$, as a function of angles, 
\begin{eqnarray*}
  P_T(\theta_1,\theta_2) 
&=& \int_{-\infty}^{\infty}\int_{-\infty}^{\infty} \ud x_1\ud x_2 |\psi(x_1,\theta_1,x_2,\theta_2)|^2\\
&=& \sum_{l,l'}^4\sum_{i,j,k}^4\sum_{i',j',k'}^4S^{(T)}_{ijk,i'j'k'}c_{ijkl}c_{i'j'k'l'}\varphi_l(\theta_1,\theta_2)\varphi_{l'}(\theta_1,\theta_2),
\end{eqnarray*}
has been shown in the left column of \fig{wavefunction}a-c for
$r=0.05a_B^*$, $r=0.10a_B^*$ and $r=0.25a_B^*$.  
The probability distribution gives the probability
$P_T(\theta_1,\theta_2)\ud\theta_1\ud\theta_2$ of finding the electrons
with angles between $\theta_1$ and $\theta_1+ \ud \theta_1$, and,
$\theta_2$ and $\theta_2+ \ud \theta_2$ from the hole, respectively. 
It is normalised such that integration over both angles from $-\pi$
to $\pi$ yields 1. It is seen that the probability distribution tends to
be more delocalised as $r$ tends to zero. This is also what we would
expect since a purely one-dimensional system has no angular
dependence. Further it is noticed that the distribution has
a butterfly shape as is seen in other cases \cite{troelsfr2009,
  citeulike:3870968}. 
In \fig{wavefunction}d we have plotted the exciton probability
distribution as function of the angle for $r$ between $0.05a_B^*$ and $r=0.25a_B^*$. The exciton probability
distribution is given by 
\begin{eqnarray*}
  P_X(\theta) 
&=& \int_{-\infty}^{\infty}\ud x |\psi_X(x,\theta)|^2 =
\sum_{j,j'}^2\sum_{i,i'}^5c_{ij}c_{i'j'}\sqrt{\frac{\pi}{\alpha_i+\alpha_{i'}}}\varphi_j^{(X)}(\theta)\varphi_{j'}^{(X)}(\theta),
\end{eqnarray*}
where $\varphi_1^{(X)}(\theta)=1/\sqrt{2\pi r}$ and
$\varphi_2^{(X)}(\theta)=|\sin\frac{\theta}{2}|/\sqrt{2\pi r}$. This
plot 
confirms the contracting behaviour of the trion probability
distribution seen in a-c. In \fig{wavefunction}e, the Hartree-Fock
probability distribution has been plotted for $r=0.10a_B^*$.
Clearly, the effects of approximating the electron-electron 
repulsion are largest along the diagonal $\theta_1=\theta_2$. 
This is emphasised in the difference plot 
\fig{wavefunction}f,
where the difference between the Hartree-Fock distribution and the full
distribution has been 
plotted in percent. The errors in the Hartree-Fock distribution vary
from -2\%  up to 4\%. 
The delocalization of the trion wave function is enhanced by the nearly equal electron and hole masses. Hence,
the centre of mass will not coincide with the position of any of the constituents but, rather, 
lie somewhere in between. This is another notable difference between trions and He atoms, for which 
the centre of mass will practically fall on top of the nucleus. As a 
consequence, both electrons and holes are almost completely delocalized around the circumference in
the realistic trion case $\sigma\approx 1.0$ and $r\approx 0.1a_B^*$, as illustrated in \fig{wavefunction}b.

Whether trions in CNTs will be detectable at room temperature is
determined by the magnitude of the trion binding energy $E_B$
relative to the thermal energy $k_BT$.
The trion binding energy depends on
the static dielectric constant $\varepsilon$ as well as the reduced
electron-hole pair mass $\mu$, through the effective Rydberg $Ry^*= 13.6
\mbox{eV}\cdot \mu/\varepsilon^2$ and the effective Bohr radius $a_B^*=
0.529\mbox{Å}\cdot \varepsilon/\mu$ \cite{citeulike:3870975,
  troelsfr2009}. The reduced mass $\mu$ is obtained from
the CNT band structure, and in order to determine the hole and electron
masses in CNTs, for a particular chiral index $(n,m)$, we used a 
non-orthogonal 
nearest neighbour 
tight binding model.
The transfer integral was chosen as\cite{citeulike:4000480} $t=-2.89 \mbox{eV}$ 
and the overlap as $s=0.1$. The soundness of these parameters follows from the fact that 
the predicted Fermi velocity of graphene $v_F=9.6\cdot 10^{5}\mbox{m/s}$ is
within 5.7\% of the average experimental value\cite{citeulike:3581542}. 
Since the CNT band structures are derived from the graphene band structure this indicates 
that effective hole and electron masses will be estimated with a similar error. 
Using this tight binding model the reduced mass was found as $\mu =
1/(m_h^{-1}+m_e^{-1})$.
The dielectric constant $\varepsilon$ on the
other hand, is determined partly by the CNTs and partly by their surroundings. 
It has been shown in several articles \cite{citeulike:7132246,citeulike:7132240,citeulike:7132241} 
that the contribution 
from the surroundings plays a significant role in the determination of
optical transition energy in CNTs. Further, the electro-static
potential $\phi$, in an quasi one-dimensional nano structure confined along $y$ and $z$, will be
governed by the perpendicular component $\varepsilon_{\perp}$, 
since \cite{landau} $\phi \propto
(\varepsilon_z\varepsilon_y x^2+\varepsilon_x\varepsilon_z
y^2+\varepsilon_x\varepsilon_y z^2 )^{-1/2}\approx \varepsilon_{\perp}^{-1}|x|^{-1}$
if the tube radius is small. It has been shown that
$\varepsilon_{\perp}$, is small for a wide range of nanotubes in
Ref. \onlinecite{citeulike:7132236}, and therefore the surrounding material is expected to
dominate the static dielectric constant $\varepsilon$.
Moreover, the results expressed in effective units are 
universal results and do not depend on $\varepsilon$. The
dependence arises only through the conversion
into physical units (eV and \AA{}) via the effective Rydberg $Ry^*$
and Bohr radius $a_B^*$,
similarly to the exciton \cite{citeulike:3870975} and
biexciton \cite{tkhp2005}.  
Since the dielectric constant depends on the material in which the CNTs
are embedded, $a_B^*$ and $Ry^*$ should be
determined for the individual experiment. Therefore $\varepsilon$
can be regarded as an experimental parameter varying from sample to
sample. 
Well-established methods for
synthesising samples with high concentrations of (6,5) CNTs exist
\cite{citeulike:6482143,citeulike:6482167} and we will therefore use
(6,5) as an example in the following. CNTs having chiral index (6,5)
have in some cases been suspended in sodium cholate
\cite{citeulike:6482167} and, in other cases, embedded in a polymeric
matrix \cite{citeulike:6482143}. The dielectric constant
$\varepsilon$ should be chosen according to either of these
materials. 
However, no exact values of $\varepsilon$ were found for the above
suspension materials. It has been shown that the photonic transition
energy as a function of environmental dielectric constant saturates
near \cite{citeulike:7132240} $\varepsilon_{env}^{(sat)} \approx
5.0$. Expecting that the surrounding material will have a dielectric
constant somewhat above the dielectric constant of air,
$\varepsilon=3.5$ is considered as a reasonable average.
The radius for (6,5) CNTs is $r=3.73$\AA{} and the effective masses of % 3870975
electron and hole turn out to be $m_e = 0.0803$ and $m_h = 0.0866$,
respectively, which leads to a reduced mass $\mu = 0.0417$. The
effective Rydberg and Bohr radius are found to be $Ry^*=0.0462$eV and
$a_B^*=44.5$\AA{}, respectively. With these values the effective radius
is $r = 0.084a_B^*$, and using this, the negative trion binding energy can
be calculated to $E_B = 1.28 Ry^* \approx 59 \mbox{meV}$.  
Hence, for this species $E_B$ is larger than $k_BT = 26\mbox{meV}$ 
and the trion state for (6,5) CNTs is expected to be detectable at 
room temperature. \\ 

In the hope that these results will stimulate measurements of the
trion binding, we have calculated the energies at $\varepsilon\in [2.0;5.0]$, which
corresponds to finding the binding energy for (6,5) CNTs suspended in
a large variety of solutions. We found that the binding energy, in this
range for $\varepsilon$, will be at least $E_B=36\mbox{meV}$ and at
most $E_B=132\mbox{meV}$.
\begin{figure}
  \centering
  \includegraphics[width=0.75\textwidth]{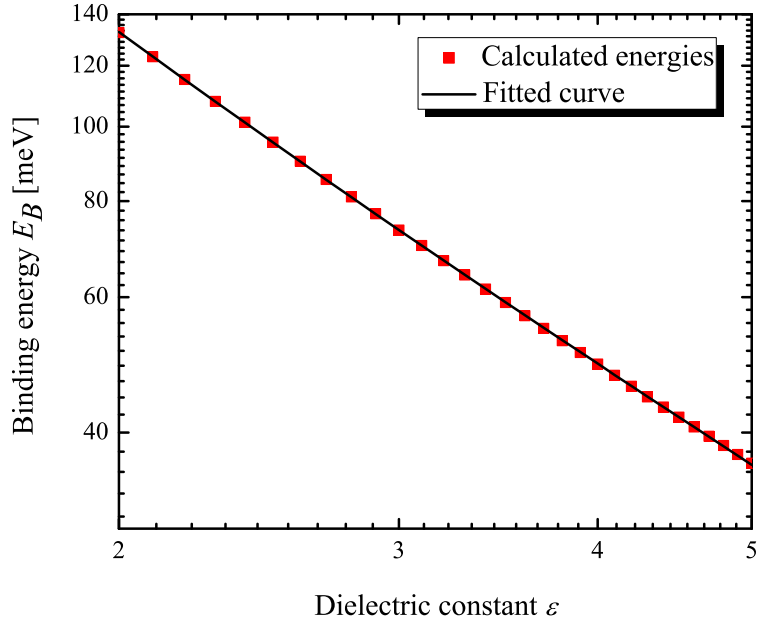}  
  \caption{Binding energy of negative trions in (6,5) CNTs as a function
    of dielectric constant $\varepsilon$.}
  \label{fig:fit}
\end{figure}
The result has been shown in \fig{fit} in a double logarithmic plot and
we also include a curve fit given by 
\begin{equation}
  E_B(\varepsilon) \approx (0.372\varepsilon^{-1.56}+0.00608) \mbox{meV}.
\end{equation}
The result is not surprising as it simply emphasises the importance of
choice of suspension material with respect to measuring the binding
energy of trions. As a consequence, experiments should be conducted with
suspension materials with low dielectric constants.
Also, the power dependence is what would be expected as a 
similar result was found for the exciton \cite{citeulike:3870975}, where
it was shown that $E_X(r) \propto r^{-0.6}Ry^*$, with $r$ in units of
$a_B^*$. Using the effective Rydberg and Bohr radius 
\begin{figure}
  \centering
  \includegraphics[width=0.75\textwidth]{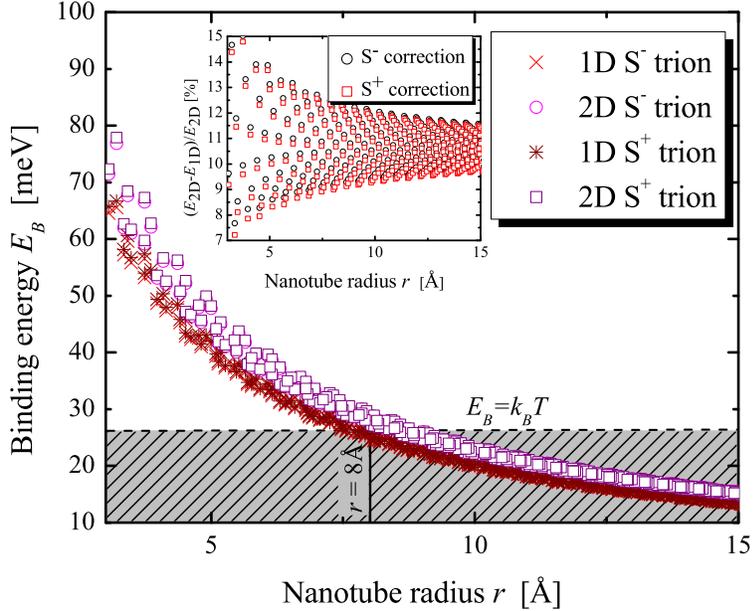}
  \caption{Binding energy of trions in CNTs in physical units using 
  effective electron and hole masses derived from a non-orthogonal tight binding scheme. 
  The shaded area illustrates the instability region for which $E_B<k_BT$. Inset: percent-wise
    difference between the one- and two-dimensional models.}
  \label{fig:real_plus_minus}
\end{figure}
it is easily shown that the exciton energy for CNTs follows a similar
power law $E_X(\varepsilon) \propto \varepsilon^{-1.4}$.\\

Keeping $\varepsilon=3.5$ constant, we found the positive and negative
trion binding energies for all semiconducting CNTs with
$3\mbox{\AA{}}\leq r \leq 15\mbox{\AA{}}$. The results are seen in
\fig{real_plus_minus}. The binding energies were found using both the
one- and two-dimensional models. 
As seen in the inset, the two-dimensional model has improved the energy,
compared with the one-dimensional one, with up to 15\%, and on average
11\%. It is noticed that in order to observe trions in CNTs at
room temperature, the radius of the CNTs must be below approximately
$8$\AA{}. Since the binding energy increases with
decreasing radius, CNTs with low radius constitute better candidates
for observing trions. Hence, CNTs with chiral index (6,5)
are very promising candidates for measuring trion binding energies.

\section{Conclusion}
\label{sec:conclussion}
In this work, trions in CNTs have been modeled as three-particle
complexes bound to the surface of a cylinder. 
We have shown that the angular behaviour plays a significant role with a
contribution of 13\% to the binding energy for cylinders with radius
$r=0.10a_B^*$. It is concluded that the energy for the $S_{0.0}^-$ trion
constitutes a fairly good approximation to both positive and negative
singlet trions with mass fraction above $0.80$. We have demonstrated
that the Hartree-Fock method applied to the $S_{0.0}^-$ trion equation
yields results that at best, are 60\% of the correct result. We conclude
that the binding energies of trions in CNTs are lowered with 11\% on
average by including the angular part of the wave function for
trions. Finally, trions are expected to be detectable in doped CNTs with
$r<8$\AA{} and we consider CNTs with chiral index $(6,5)$ as a very
good candidate for measuring trion binding energies.

%Merlin.mbs v4.21 2009-07-09.
%

%\bibliography{draft.bib}
\end{document}